\documentclass[11pt,preprint,preprintnumbers,nofootinbib,
groupedaddress,superscriptaddress,amsmath,amssymb]{revtex4}
\pdfoutput=1 

\usepackage{amsfonts}
\usepackage{graphics}
\usepackage{epsfig}
\usepackage{url}
\usepackage{multirow}
\usepackage{feynmp}
\newcommand {\be}{\begin{equation}}
\newcommand {\ee}{\end{equation}}
\newcommand {\ba}{\begin{eqnarray}}
\newcommand {\ea}{\end{eqnarray}}

\newcommand {\tanb}{$\tan\beta~$}
\newcommand {\ra}{\rightarrow}

\begin{document} 

\title{\boldmath Off-diagonal Yukawa Couplings in the $s$-channel Charged Higgs Production at LHC}

\author{Majid Hashemi}
\email{hashemi_mj@shirazu.ac.ir}
\author{Seyyed Mohammad Zebarjad}
\author{Hossein Bakhshalizadeh}

\affiliation{Physics Department and Biruni Observatory,College of Sciences, Shiraz University, Shiraz 71454, Iran}


\begin{abstract}
The search for the heavy charged Higgs ($m_{H^{\pm}}>m_{\textnormal{top}}$) has been mainly based on the off-shell top pair production process. However, resonance production in $s$-channel single top events is an important channel to search for this particle. In a previous work, it was shown that this process, i.e., $qq' \ra H^{+} \ra t\bar{b}$ + h.c., can lead to comparable results to what is already obtained from LHC searches. What was obtained was, however, based on diagonal Yukawa couplings between incoming quarks assuming $c\bar{s}$ as the main incoming pair due to the CKM matrix element being close to unity. The aim of this paper is to show that off-diagonal couplings, like $c\bar{b}$, may lead to substantial contributions to the cross section, even if the corresponding CKM matrix element is two orders of magnitude smaller. For this reason, the cross section is calculated for each initial state including all diagonal and off-diagonal terms, and all is finally added together to get the total cross section which is observed to be $\sim 2.7$ times larger than what is obtained from $c\bar{s}$ initial state. Results are eventually reflected into 95$\%$ C.L. exclusion and 5$\sigma$ discovery contours at different integrated luminosities of LHC. A reasonable coverage of the parameter space is obtained by the 95$\%$ C.L. exclusion contour.
\end{abstract}

\maketitle

\section{Introduction}
The search for the charged Higgs boson within Supersymmetric Standard Model (MSSM) at the Large Hadron Collider (LHC) is currently extending the excluded area in the parameter space ($m_{H^{\pm}},$\tanb) with no evidence of the particle in the low mass area ($m_{H^{\pm}}<m_{\textnormal{top}}$). Here, \tanb is the ratio of vacuum expectation values of the two Higgs doublets. \\
The current main limits on the mass of the charged Higgs come from LEP II direct and indirect searches which, all together, set a lower limit on the charged Higgs mass as $m_{H^{\pm}}>125~ \textnormal{GeV}$ \cite{lepexclusion1,lepexclusion2}. The Tevatron searches by the D0 \cite{d01,d02,d03,d04} and CDF Collaborations \cite{cdf1,cdf2,cdf3,cdf4} exclude high \tanb values, however, they are confirmed and extended by current LHC results \cite{cmsindirect,atlasindirect,atlasdirect,atlasdirect2,cmsdirect,cmsdirect2}.\\
There are also stringent constraints on the charged Higgs mass from flavor physics studies. A study of recently completed BaBar data analysis \cite{babar} based on $b \ra s\gamma$ excludes $m_{H^{\pm}}<380$ GeV independent of \tanb \cite{bsg}. The LHC searches are nevertheless important and will provide, in case, an independent confirm of the exclusion regions from flavor bounds. Therefore in this paper, we rely on the direct search results from LHC which exclude \tanb$>$ 50 in the heavy charged Higgs area, i.e., $m_{H^{\pm}}>200$ GeV \cite{atlasdirect2}. The presented results, however, are extended to $m_{H^{\pm}}=400$ GeV which is beyond the current flavor physics limits.\\
The ongoing analyses at LHC focus on $g\bar{b} \ra t H^{-}$ process to search for the heavy charged Higgs. However, single top events have also been proved to be a significant source of the charged Higgs in both low and high mass regions. The light charged Higgs may be produced in a $t$-channel single top production through the top quark decay \cite{st1}. The heavy charged Higgs is, however, produced directly through the $s$-channel single top production with the signature of such events being the kinematic differences from pure Standard Model (SM) events \cite{st2}. The analysis performed in \cite{st2} has led to promising results comparable to what has been obtained from the analysis of $g\bar{b} \ra t H^{-}$ process at LHC  \cite{c1,c2,a1,a2}\\
The aim of this paper is to show that there is still room to improve the signal sensitivity in $s$-channel single top analysis \cite{st2}. An analysis of the parton distribution functions (PDF) inside the incoming protons shows that heavy quark PDF ($c$ and $b$ quarks) is not negligible and can lead to sizable contribution to the rate of incoming partons in the interaction. On the other hand the vertex coupling in a process like $c\bar{b} \ra H^+$ is proportional to the $b$-quark mass at high \tanb while the corresponding diagonal coupling which appears at $c\bar{s} \ra H^+$ interaction is proportional to the $s$-quark mass. One should of course take into account the CKM matrix element suppression in the former, however, as will be shown, the suppression is not strong enough to decrease the rate dramatically. In fact it turns out that $c\bar{b}$ initiated process has a larger cross section than that initiated from $c\bar{s}$.\\
In what follows, details of the $s$-channel single top cross section calculation is presented, with a list of quark masses and CKM matrix element values used in the analysis. All possible initial states are included in the calculation and a total cross section is obtained as the sum of diagonal and off-diagonal couplings between incoming quarks and compared with what is obtained from the main diagonal coupling $c\bar{s}$. In order to visualize the results, event selection efficiencies from \cite{st2} are used and updated contours are presented for a 5$\sigma$ discovery or 95$\%$ C.L. exclusion.    
  
\section{Cross Section of Heavy Charged Higgs Production in Single Top Events}
\begin{figure}
 \centering
 \includegraphics[width=0.6\textwidth]{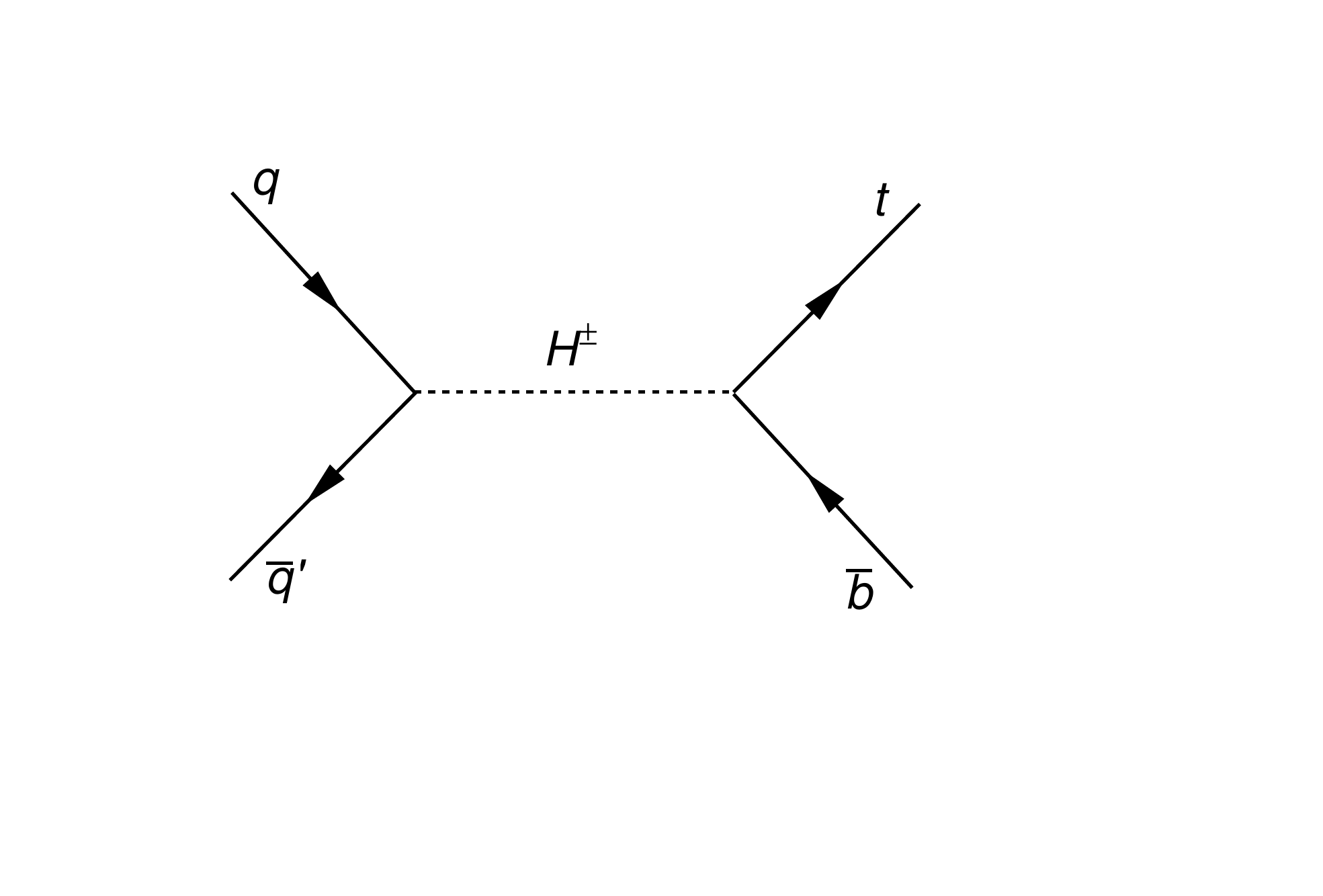}
 \caption{The signal production process. The hermitian conjugate of the above process is included throughout the paper in all calculations even if not explicitly stated. \label{diagram}}
 \end{figure}

In this section a description of cross section calculation based on Yukawa couplings is presented. The analysis is based on MSSM, $m_{h}$-max scenario \cite{lepexclusion2} with the following parameters: $M_{2}=200$ GeV, $M_{\tilde{g}}=800$ GeV, $\mu=200$ GeV and $M_{SUSY}=1$ TeV. In order to be more specific, the Feynman diagram under study is illustrated in Fig. \ref{diagram}. The final evaluation and parameter dependence of the cross section is obtained using CompHEP \cite{comphep,comphep2}. To this end, vertex couplings (to be described in the next sub-section) are implemented in CompHEP and the cross section is calculated using Monte Carlo approach requiring a statistical error less than a percent.

\subsection{Yukawa Coupling Lagrangian}
\begin{figure}
 \centering
 \includegraphics[width=0.6\textwidth]{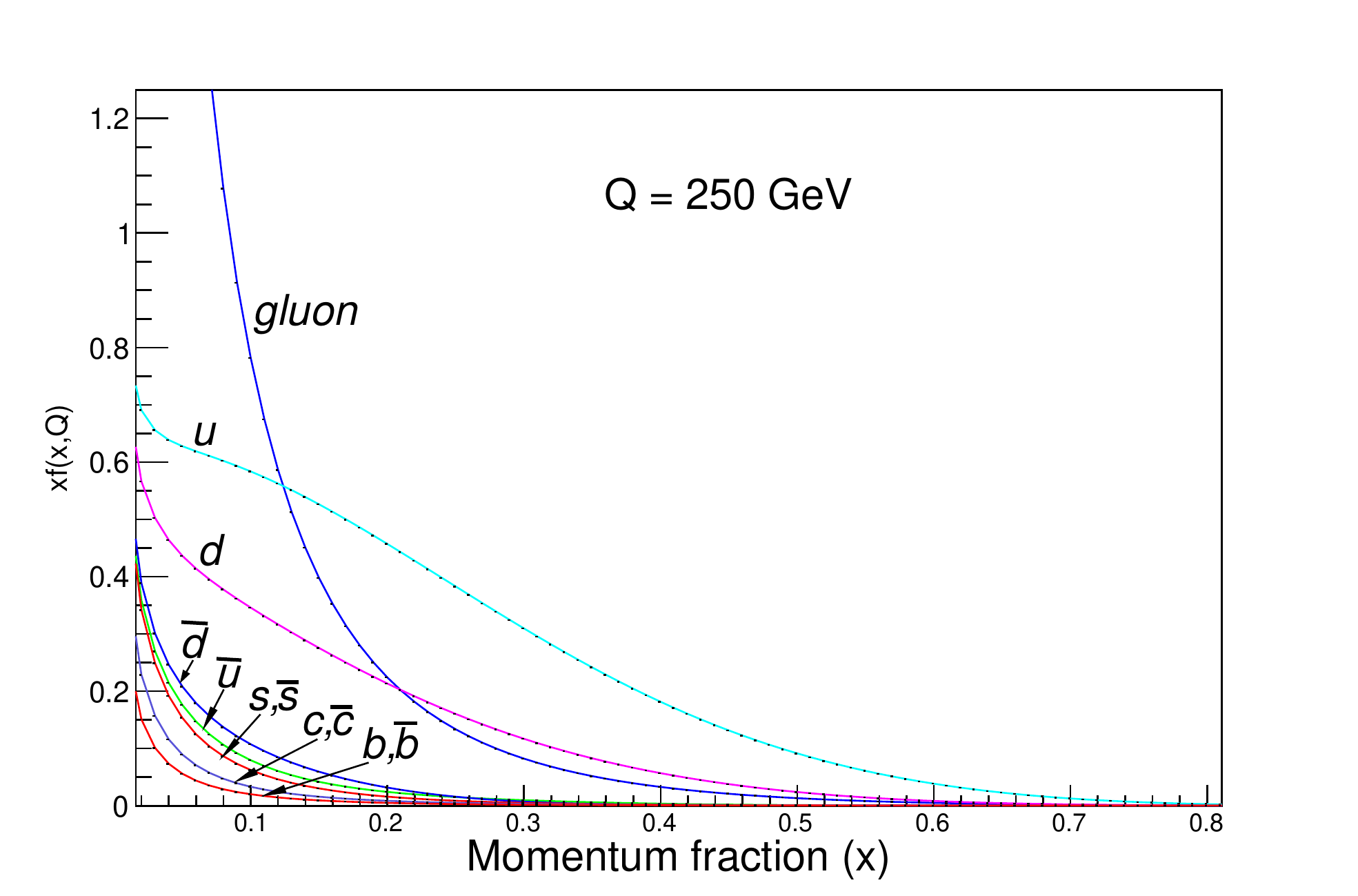}
 \caption{The parton distribution functions at $Q=250$ GeV, using CTEQ 6.6. \label{pdfs}}
 \end{figure}

The charged Higgs interaction with leptons and quarks can be formulated with the following Lagrangian:
\be
\mathcal{L}=\sqrt{2\sqrt{2}G_F}~H^+\left[ V_{UD}(m_U \cot \beta ~\bar{U} P_L D ~+~ m_D \tan \beta ~\bar{U} P_R D ) + m_l \tan\beta ~\bar{\nu} P_R l \right] 
\label{lagr}
\ee 
where $P_L~(P_R)$ are left (right) hand projection operators, $V_{UD}$ is the CKM matrix element, and an implicit sum over $U$ (up type quarks) and $D$ (down type quarks) is assumed. The last term is not under consideration here, but the first two terms describe charged Higgs interaction with quarks. The interaction, as is seen from Eq. \ref{lagr}, is sensitive to the quark masses as well as the CKM matrix elements. This Lagrangian can be used to calculate a parton level cross section, however, in real proton-proton interactions, another issue is how likely a parton of a given type comes out of the proton and takes part in the interaction. This effect is described by the parton distribution function used in the analysis. Figure \ref{pdfs} shows typical distributions of the partons in a proton at a negative four momentum transfer set to $Q=250$ GeV (a charged Higgs with $m_{H^{\pm}}=250$ GeV). As is seen from Fig. \ref{pdfs} heavy quarks (even the $b$-quark) can still be visible although with a much smaller probability than valence quarks and gluons. A $b$-quark may appear directly or through a gluon splitting and its contribution to the Yukawa coupling is proportional to its mass at high \tanb. Therefore it may not be surprising that a $c\bar{b}$ initial state makes a larger contribution than diagonal $c\bar{s}$ term in spite of the strong CKM suppression.

\subsection{Quark Masses and CKM Quark-Mixing Matrix Elements}
For our calculations, we use Particle Data Group (pdg) data \cite{pdg} for quark masses as well as the CKM matrix elements as listed in Tabs. \ref{qm} and \ref{ckm} respectively. The CKM elements listed in Tab. \ref{ckm} are those related to the initial state quarks (incoming partons). The final state is set to a pair of top and bottom quark. Therefore the only CKM matrix element for the final state is $V_{tb}$ which is assumed to be unity (The recent CMS measurement at 7 TeV implies $V_{tb}=1.14\pm0.22$).\\
It should be noted that the presented quark masses in Tab. \ref{qm} are used for performing their running to the proper scale of the interaction which is the charged Higgs mass. The running process is performed by FeynHiggs 2.8.3 \cite{feynhiggs} which is linked to CompHEP for cross section calculation. This process is repeated for each charged Higgs mass hypothesis and is done for all quarks involved in the parton level interaction.\\
As an example, using FeynHiggs, the following quark masses are obtained at the scale of $m_{H^{\pm}}=200$ GeV: $m_c=0.57$ GeV, $m_s=0.05$ GeV, $m_b=2.63$ GeV, $m_t=169$ GeV. These values result in $\Gamma_{cs}=2.43\times10^{-3}$ GeV, $\Gamma_{tb}=0.56$ GeV with \tanb= 50, where $\Gamma_{cs}(\Gamma_{tb})$ is charged Higgs decay rate to a pair of $c\bar{s}$($t\bar{b}$). Using a tree level calculation based on the Yukawa coupling Lagrangian Eq. \ref{lagr}, the partial rate of the charged Higgs decay to quark pairs is obtained as in Eq. \ref{gamma},
\be
\Gamma_{H^{\pm}\ra U\bar{D}}=\frac{3\sqrt{2}G_F V^{2}_{UD}}{8\pi}m_{H^{\pm}} \left (1-\frac{m_U^2}{m_{H^{\pm}}^{2}}\right )^{2} \left [ m_U^2 \cot^2\beta+m_D^2\tan^2\beta  \right ]
\label{gamma}
\ee
where $U$ and $D$ denote up-type and down-type quarks. Inserting quark masses mentioned above in Eq. \ref{gamma} the following partial decay rates are obtained : $\Gamma_{cs}=2.45\times10^{-3}$ GeV, $\Gamma_{tb}=0.55$ GeV. These values are based on quark masses at the charged Higgs mass scale and are in reasonable agreement with FeynHiggs results which are used by CompHEP for cross section calculation.
\subsection{Cross section Calculation}
\begin{figure}
 \centering
 \includegraphics[width=0.6\textwidth]{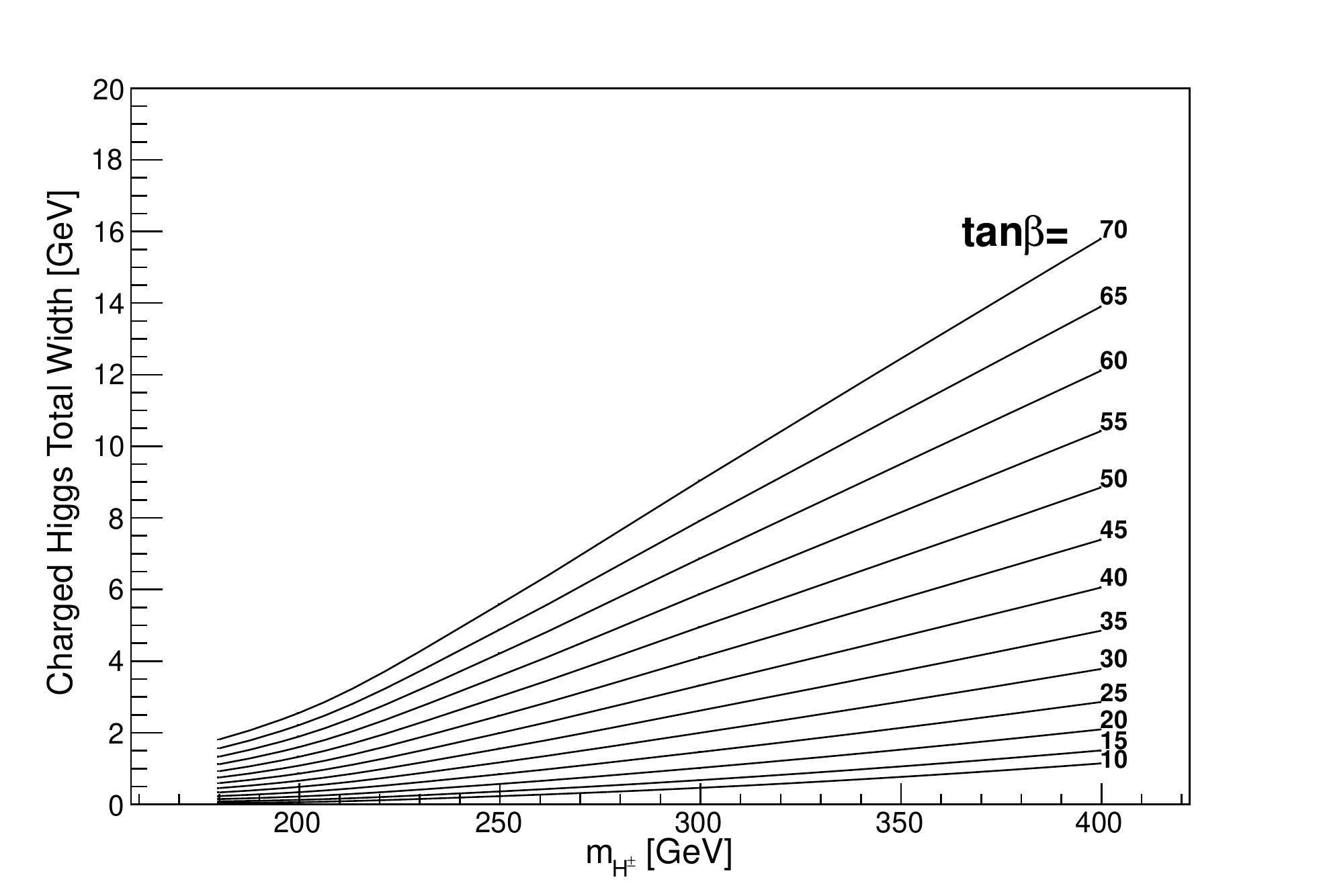}
 \caption{Charged Higgs total width at different \tanb. \label{widths}}
\end{figure}

The main approach for cross section calculation in this paper is based on using CompHEP. The charged Higgs total width is calculated for several charged Higgs mass and \tanb values using FeynHiggs. Results are shown in Fig. \ref{widths} and used by CompHEP for calculating cross sections. The integration over parton level cross sections is performed in CompHEP using CTEQ 6.6 parton distribution function (PDF) to obtain the cross section at real proton-proton interactions at LHC nominal center of mass energy $\sqrt{s}=14$ TeV.\\
In order to check cross section values, a second approach is also adopted by calculating parton level cross sections as in Eq. \ref{sigmahat},
\be
\hat{\sigma}=\frac{2j+1}{(2s_1+1)(2s_2+1)}\frac{16\pi m^2_{H^{\pm}}}{\hat{s}}\frac{\Gamma(H^{\pm}\rightarrow c\bar{s})\Gamma(H^{\pm}\rightarrow t\bar{b})}{(\hat{s}-m^2_{H^{\pm}})^2+m^2_{H^{\pm}}\Gamma^2_{\textnormal{total}}}
\label{sigmahat}
\ee
where $\hat{s}=x_i x_j s$, $s$ is the square of the center of mass energy ($\sqrt{s}=14$ TeV) and $x_i$ and $x_j$ are proton momentum fractions carried by the two incoming partons. The spin factor in Eq. \ref{sigmahat} uses $j=0$ for the charged Higgs spin and $s_1=s_2=1/2$ for the spin of incoming partons. The proton-proton cross section is then obtained by inserting $\hat{\sigma}$ (Eq. \ref{sigmahat}) in Eq. \ref{sigmatot} which takes into account the parton distribution functions $f(x_i,Q,i)$,
\be
\sigma=\sum_{i,j}\int dx_i \int dx_j ~f(x_i,Q,i) f(x_j,Q,j) ~\hat{\sigma}
\label{sigmatot}
\ee   
In Eq. \ref{sigmatot}, $i$ and $j$ denote the parton flavor and $Q$ (negative momentum transfer) is taken as the charged Higgs mass. The cross section calculation using Eq. \ref{sigmatot} requires parton distributin functions, $f(x_i,Q,i)$, which are accessible through LHAPDF. In our calculation LHAPDF 5.8.6 \cite{lhapdf} is used for this purpose and a code is written to numerically integrate over $x_i$ and $x_j$ and calculate the cross section. Results from CompHEP and our calculation for $m_{H^{\pm}}=200$ GeV and \tanb= 50 are $\sigma(c\bar{s}\ra H^+ \ra t\bar{b})_{\textnormal{(CompHEP)}}=1.46 ~pb$ and $\sigma(c\bar{s}\ra H^+ \ra t\bar{b})_{\textnormal{(Our calculation)}}=1.40 ~pb$. Therefore cross sections are calculated correctly based on quark masses at the proper scales.\\ 

\begin{table}[h]
\begin{center}
\begin{tabular}{cc}
\hline
Quark flavor & Mass [GeV] \\
\hline
$u$ & 0.0023\\
\hline
$d$ & 0.0048\\
\hline
$s$ & 0.095\\
\hline
$c$ & 1.275 \\
\hline
$b$ & 4.18 \\
\hline
$t$ & 173 \\  
\hline
\end{tabular}
\end{center}
\caption{Quark masses according to PDG 2012 \cite{pdg}. The light quark masses are evaluated in the $\bar{MS}$ scheme at a scale $\mu \sim 2$ GeV. The cross section is almost insensitive to the light quarks. The $c$- and $b$-quark masses are the running masses in the $\bar{MS}$ scheme. These values are used as input for running to the proper scale of the interaction which is taken as the charged Higgs mass. \label{qm}}
\end{table}

\begin{table}[h]
\begin{center}
\begin{tabular}{cc}
\hline
CKM matrix element & Value \\
\hline
$V_{ud}$ & 0.97\\
\hline
$V_{us}$ & 0.22\\
\hline
$V_{ub}$ & 0.004 \\  
\hline
$V_{cd}$ & 0.23\\
\hline
$V_{cs}$ & 1.00 \\
\hline
$V_{cb}$ & 0.04 \\
\hline
\end{tabular}
\end{center}
\caption{CKM quark-mixing matrix elements according to PDG 2012 \cite{pdg}. \label{ckm}}
\end{table}
\subsection{Results}
\begin{figure}[h]
 \centering
 \includegraphics[width=0.6\textwidth]{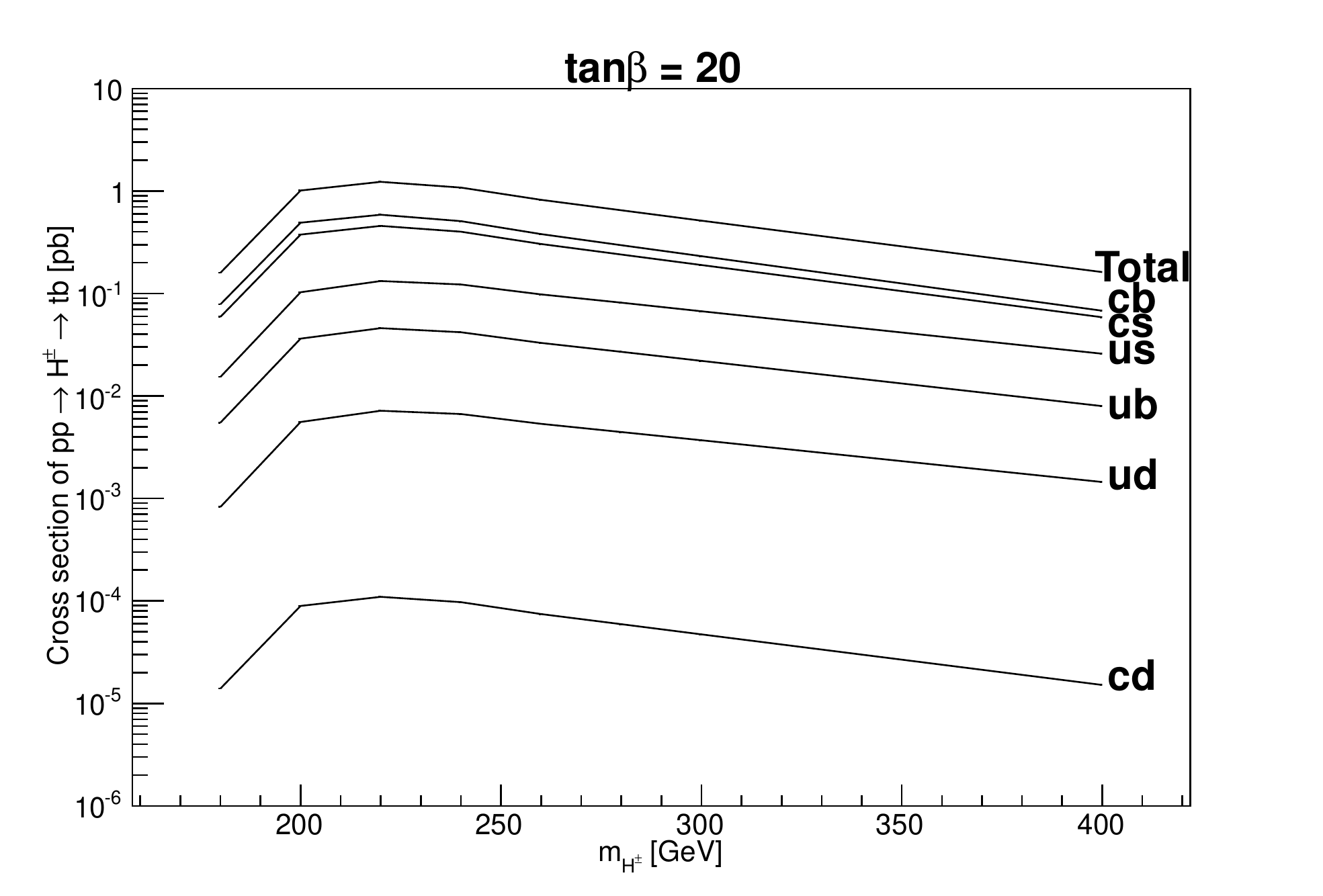}
 \caption{Cross section of $pp \ra H^{\pm} \ra tb$ at \tanb = 20. Contribution of different initial states as well as the total value are shown separately. \label{xsec20}}
\end{figure}

\begin{figure}[h]
 \centering
 \includegraphics[width=0.6\textwidth]{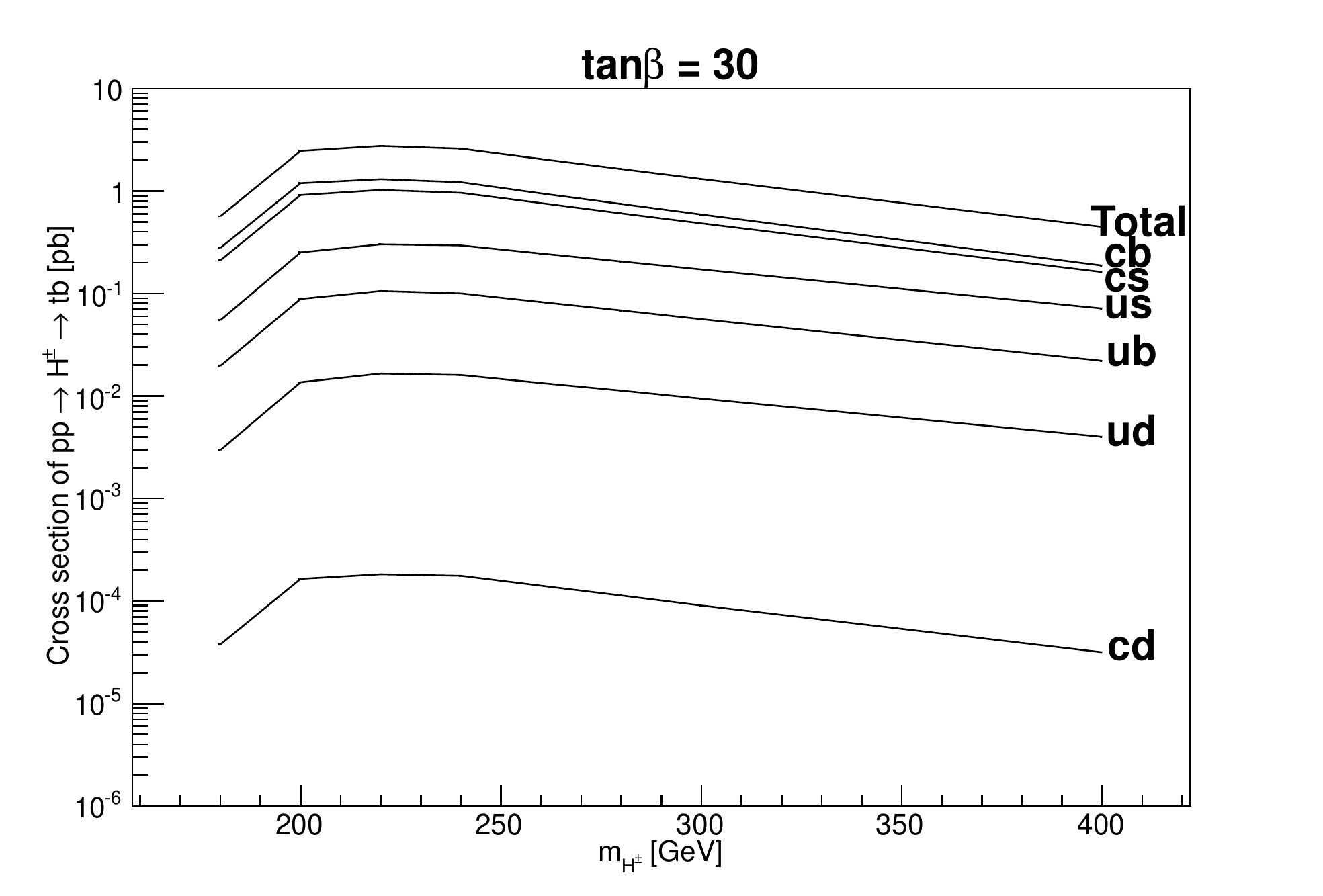}
 \caption{Cross section of $pp \ra H^{\pm} \ra tb$ at \tanb = 30. Contribution of different initial states as well as the total value are shown separately. \label{xsec30}}
\end{figure}

\begin{figure}
 \centering
 \includegraphics[width=0.6\textwidth]{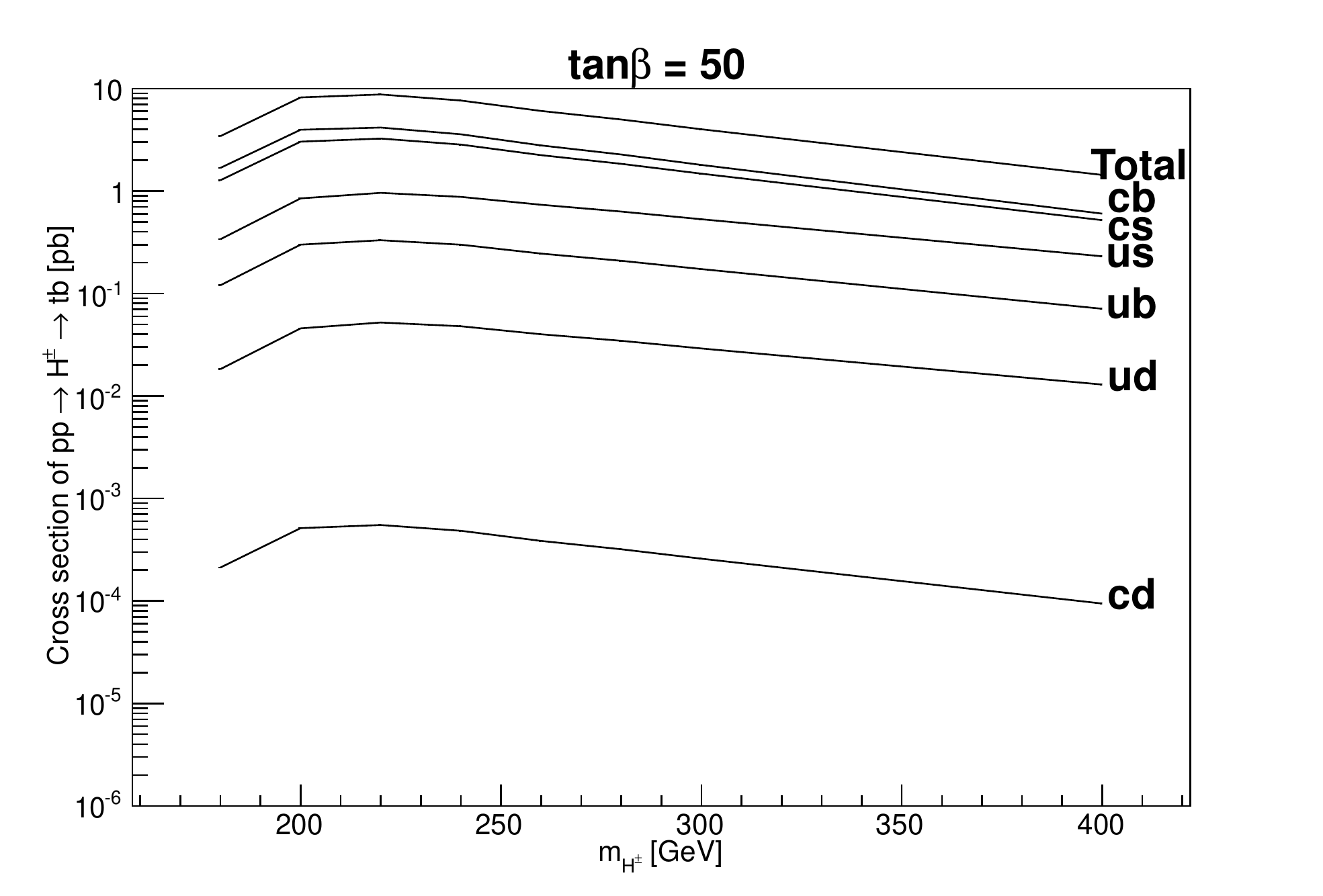}
 \caption{Cross section of $pp \ra H^{\pm} \ra tb$ at \tanb = 50. Contribution of different initial states as well as the total value are shown separately. \label{xsec50}}
\end{figure}

\begin{figure}
 \centering
 \includegraphics[width=0.6\textwidth]{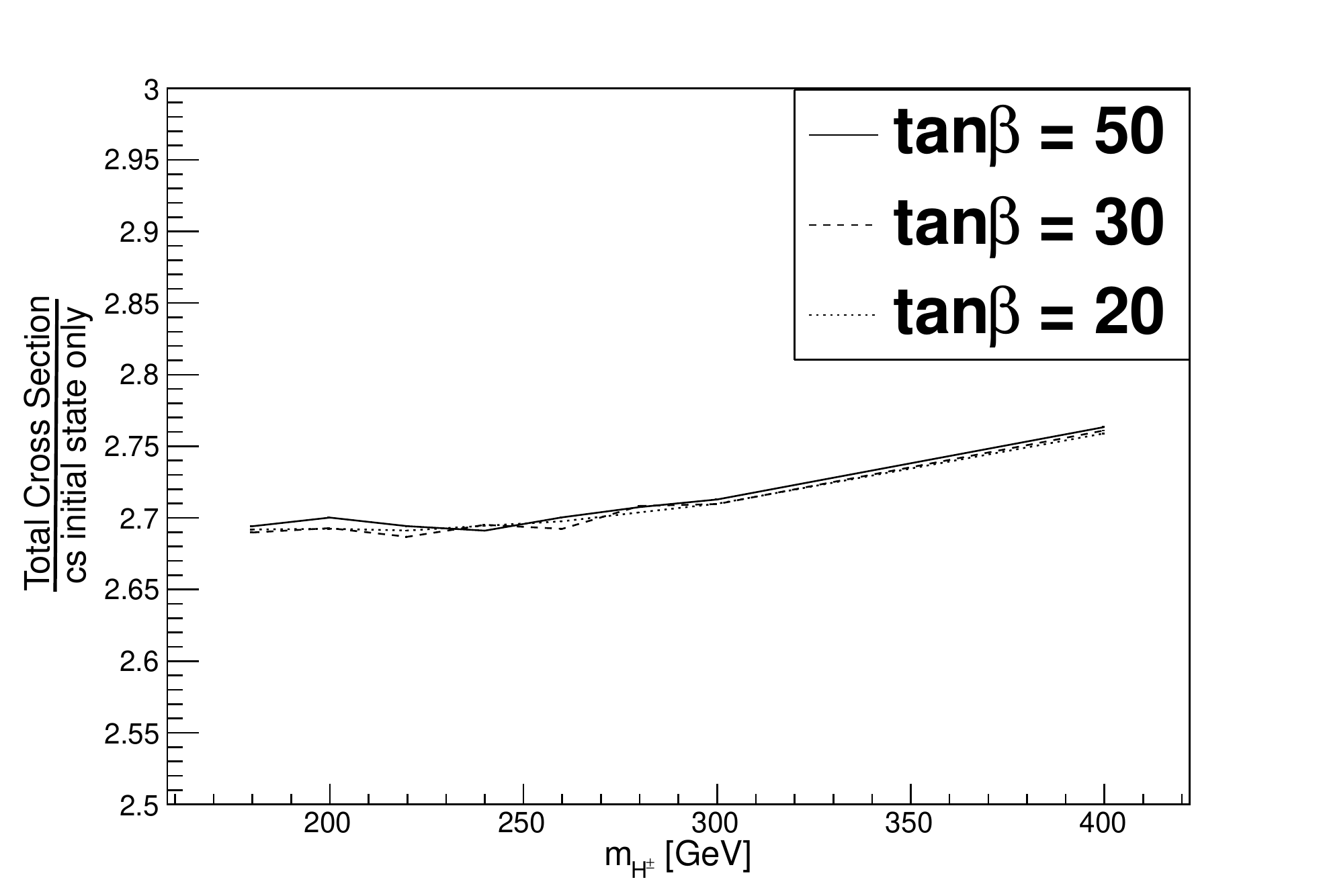}
 \caption{Ratio of total cross section of $pp \ra H^{\pm} \ra tb$ to that of only $cs$ initial state. \label{totaltocs}}
\end{figure}

\begin{figure}[h]
 \centering
 \includegraphics[width=0.6\textwidth]{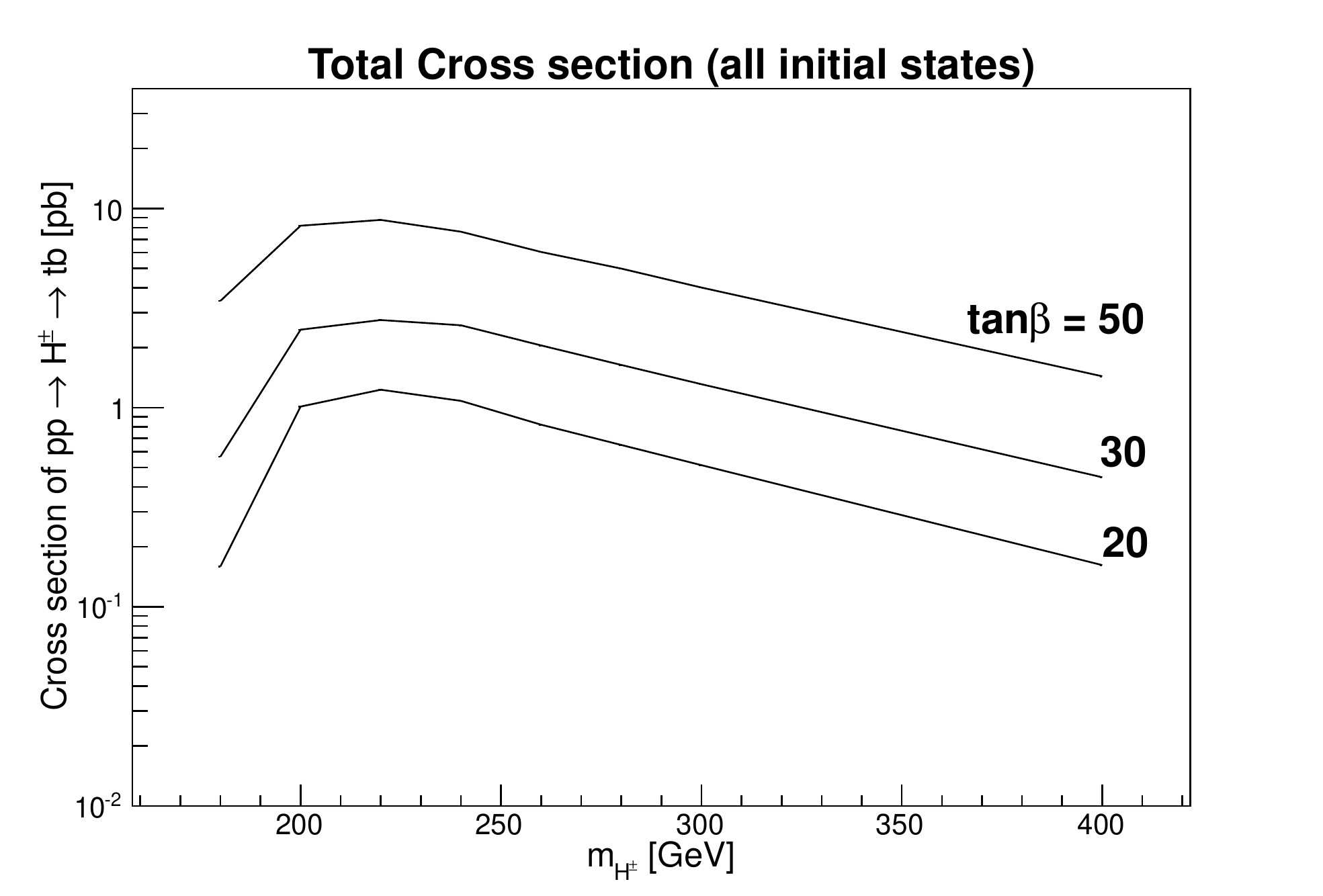}
 \caption{Cross section comparison at different \tanb. \label{xseccomparison}}
\end{figure}

In this section, results of $pp \ra H^{\pm} \ra tb$ cross section calculation are presented for different \tanb values. In Figs. \ref{xsec20}, \ref{xsec30} and \ref{xsec50} results are presented including all initial states which are shown in separated curves. The total cross section is the sum of all initial states. As is seen the $cb$ initial state has the largest contribution to the total cross section for any charged Higgs mass and \tanb. The ratio of total cross section to that of $cs$ initial state is shown in Fig. \ref{totaltocs}. Results are in agreement with \cite{offdiagonal} where a discussion on the contribution of $cb$ and $cs$ initial states has been presented. Finally Fig. \ref{xseccomparison} compares cross sections at different \tanb values including all initial states and reveals the fact that at high \tanb the cross section grows rapidly. 

\section{The 95$\%$ C.L. Exclusion and 5$\sigma$ Discovery Contours}
\begin{figure}[h]
 \centering
 \includegraphics[width=0.6\textwidth]{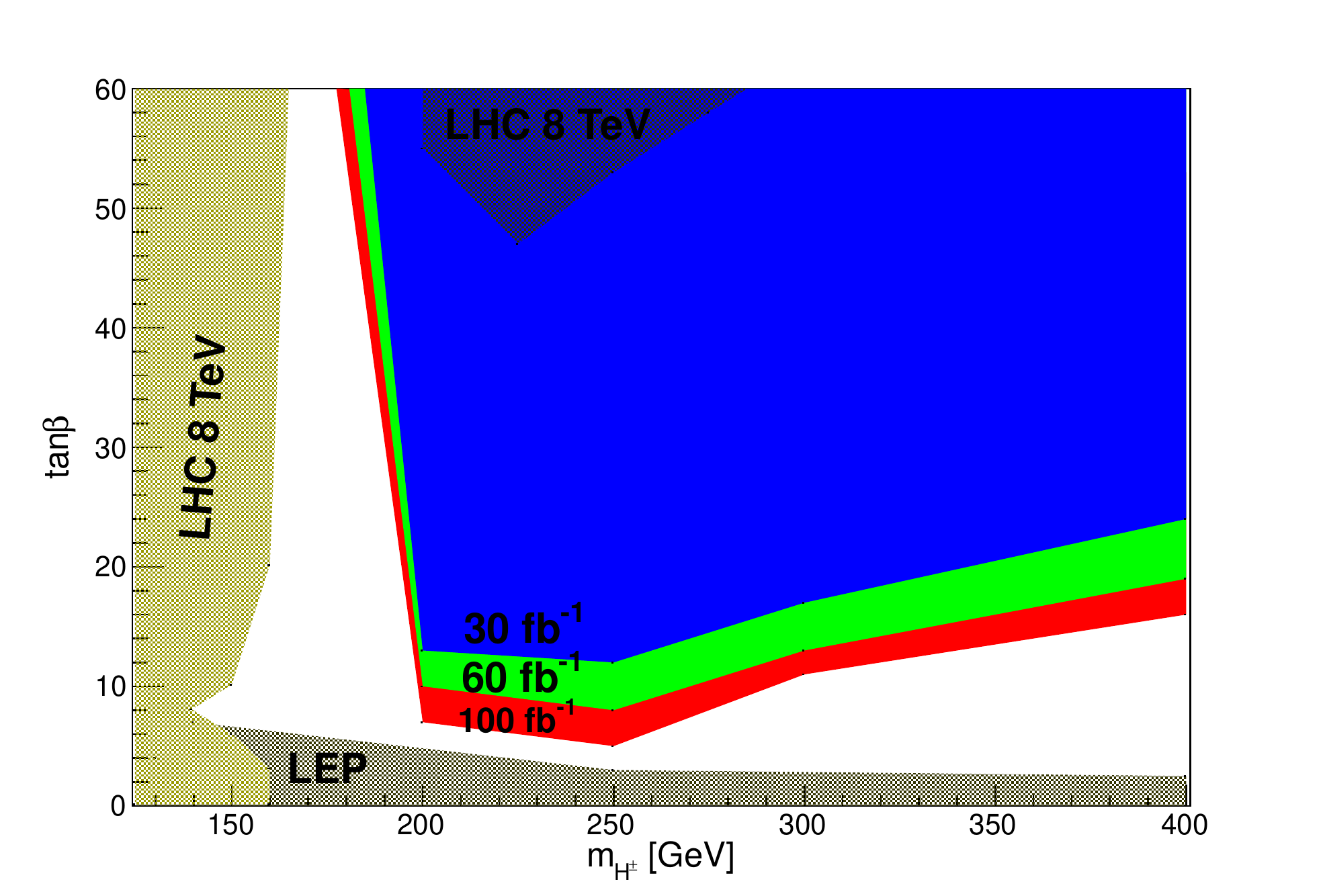}
 \caption{The 95$\%$ C.L. exclusion contour at different integrated luminosities of 30, 60 and 100 $fb^{-1}$. The regions with ``LEP'' and ``LHC'' labels refer to \cite{lepexclusion2} and \cite{atlasdirect2}. \label{95cl}}
\end{figure}

\begin{figure}[h]
 \centering
 \includegraphics[width=0.6\textwidth]{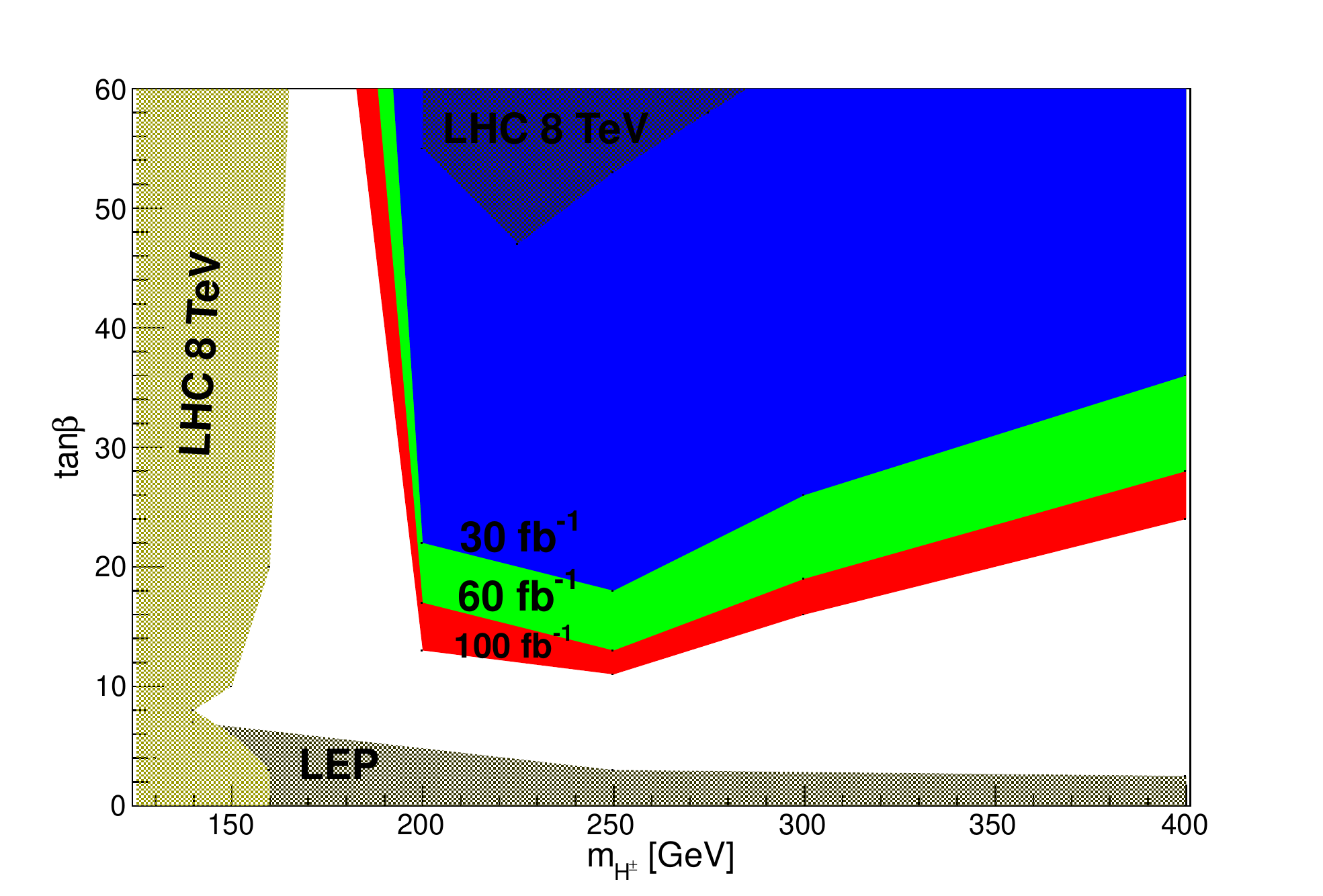}
 \caption{The 5$\sigma$ discovery contour at different integrated luminosities of 30, 60 and 100 $fb^{-1}$. The regions with ``LEP'' and ``LHC'' labels refer to \cite{lepexclusion2} and \cite{atlasdirect2}. \label{5sigma}}
\end{figure}

Results of \cite{st2} are based on a cross section calculation including only $cs$ initial state. As Fig. \ref{totaltocs} shows, the cross section is $\sim 2.7$ times larger if all initial states are included. This effect is almost independent of the charged Higgs mass and \tanb. Therefore using the same event selection efficiencies as in \cite{st2} and updated cross sections obtained in this paper, contours of 95$\%$ C.L. exclusion and 5$\sigma$ discovery are produced using TLimit code implemented in ROOT \cite{root}. Figures \ref{95cl} and \ref{5sigma} show the 95$\%$ C.L. exclusion and 5$\sigma$ discovery contours respectively. In Figs. \ref{95cl} and \ref{5sigma}, the excluded region obtained in \cite{atlasdirect2} has been shown. Small \tanb values are excluded by LEP \cite{lepexclusion2}. 

\section{Conclusions}
The $s$-channel charged Higgs production was revisited with special care on the contribution of different incoming partons in the interaction. The total cross section was obtained including all initial states and it was concluded that off-diagonal terms in the Yukawa interaction of the charged Higgs and quarks play an important role even though partially suppressed by the CKM matrix elements. The total cross section was obtained to be $\sim 2.7$ times the dominant digonal term, i.e., $cs$ initial state. Using selection efficiencies from an earlier analysis, contrours of exclusion and discovery were updated. Results show that with this channel, almost all parameter space in the mass range  $200$ GeV$<m_{H^{\pm}}<300$ GeV can be excluded even at \tanb values as low as 10. This is a result which has not yet been obtained using current LHC experiments and is worth considering in their analyses.
\section{Acknowledgement}
This work was completed using the computing cluster facility at Sciences Department of Shiraz University. We would like to appreciate Dr. Mogharrab for his efforts in maintaining the cluster operation.
\pagebreak

\end{document}